\documentclass[aps,prd,twocolumn, superscriptaddress,nofootinbib]{revtex4}
\usepackage{fancyhdr}
\usepackage{amsthm}
\usepackage{graphicx}
\usepackage{subfigure}
\usepackage{hyperref}
\usepackage{enumerate}
\usepackage{amssymb, amsmath}
\usepackage{color}

\def\beq{\begin{equation}}
\def\eeq{\end{equation}}
\def\bea{\begin{eqnarray}}
\def\eea{\end{eqnarray}}

\def\bwt{\begin{widetext}}
\def\ewt{\end{widetext}}

\begin{document}

\title{Empty Black Holes, Firewalls, and the Origin of Bekenstein-Hawking Entropy}

\author{Mehdi Saravani}\email{msaravani@pitp.ca}
\affiliation{Perimeter Institute
for Theoretical Physics, 31 Caroline St. N., Waterloo, ON, N2L 2Y5, Canada}
\affiliation{Department of Physics and Astronomy, University of Waterloo, Waterloo, ON, N2L 3G1, Canada}
\author{Niayesh Afshordi}\email{nafshordi@pitp.ca}
\affiliation{Perimeter Institute
for Theoretical Physics, 31 Caroline St. N., Waterloo, ON, N2L 2Y5, Canada}
\affiliation{Department of Physics and Astronomy, University of Waterloo, Waterloo, ON, N2L 3G1, Canada}
\author{Robert B. Mann}\email{rbmann@uwaterloo.ca}
\affiliation{Department of Physics and Astronomy, University of Waterloo, Waterloo, ON, N2L 3G1, Canada}
\affiliation{Perimeter Institute
for Theoretical Physics, 31 Caroline St. N., Waterloo, ON, N2L 2Y5, Canada}

\begin{abstract}
We propose a novel solution for the endpoint of gravitational collapse, in which spacetime ends (and is orbifolded) at a microscopic distance from black hole event horizons. This model is motivated by the emergence of singular event horizons in the gravitational aether theory, a semi-classical solution to the cosmological constant problem(s), and thus suggests a catastrophic breakdown of general relativity close to black hole event horizons. A similar picture emerges in {\it fuzzball} models of black holes in string theory, as well as the recent {\it firewall} proposal to resolve the information paradox. We then demonstrate that positing a surface fluid in thermal equilibrium with Hawking radiation, with vanishing energy density (but non-vanishing pressure) at the new boundary of spacetime, which is required by Israel junction conditions, yields a thermodynamic entropy that is identical to the Bekenstein-Hawking area law, $S_{BH}$, for charged rotating black holes. To our knowledge, this is the first derivation of black hole entropy which only employs {\it local} thermodynamics. Furthermore, a model for the microscopic degrees of freedom of the surface fluid (which constitute the micro-states of the black hole) is suggested, which has a finite, but Lorentz-violating, quantum field theory. Finally, we comment on the effects of physical boundary on Hawking radiation, and show that relaxing the assumption of equilibrium with Hawking radiation sets $S_{BH}$ as an upper limit for Black Hole entropy.
\end{abstract}
\maketitle
\flushbottom

\section{Introduction}

General relativity (GR) predicts that the endpoint of gravitational collapse of (nonrotating and neutral) matter is a (Schwarzschild) Kerr-Newman black hole. While there might be a metric singularity at the horizon of the  black hole, there is no real curvature singularity. However, there is a real curvature singularity at the centre of a black hole. Since the curvature-invariants remain small at black hole horizons, it is widely believed that black hole solutions of GR are good approximations to the ``real'' geometry of spacetime all the way to the singularity, except perhaps for a neighbourhood of the singularity, or at time-scales comparable to the Hawking evaporation time,  where/when quantum mechanical effects become important.

Nevertheless, there are a number of arguments against the validity of the semi-classical nature of GR black hole solutions inside event horizons.  If we consider GR as an effective field theory, its expected cut-off will be around the Planck energy, $M_P$.  Quantum loop corrections should therefore be suppressed by powers of ${\cal O}({\cal R}/M^2_P)$, which are negligible around horizons of macroscopic black holes. However, {\it non-perturbative} quantum effects can be big: while  the tunnelling rate is suppressed by $\exp(-S_E)\ll 1$, where $S_E$ is the Euclidean action of the instanton connecting GR solutions to other  states in the full theory of quantum gravity,  the number of such states is estimated to be $\sim \exp(S_{BH})\gg 1$, with $S_{BH}$ being the Bekenstein-Hawking  entropy of black holes. Interestingly, $S_{E} = S_{BH}$ for Euclideanized GR black holes, and thus the non-perturbative decay of the semi-classical solutions can be quite fast (i.e. much faster than the Hawking evaporation time) \cite{Mathur:2008kg,Mathur:2009zs}. While the nature of the end-state of gravitational collapse depends on the full phase space of the theory of quantum gravity, in the context of string theory, it has been argued that {\it fuzzball} solutions provide the correct multiplicity and asymptotics to represent a microscopic description of GR black hole macro-states (e.g. see \cite{Mathur:2012zp} and references therein). While fuzzball solutions approximate GR black holes at large distances, they diverge from the classical solution (or each other) at/around  the classical horizon. In particular, fuzzball solutions {\it do not} have any event horizons or singularities, although they contain ergo-regions, which could produce analogue of Hawking radiation.  Moreover, the spacetime ``ends'' at a minimal spatial area comparable to that of the classical event horizon. The latter is the most significant macroscopic difference between the fuzzballs, and their semi-classical counter-parts, which we will capture below in our construction. 

Very recently, a similar picture has emerged from a reconsideration of the black hole information paradox \cite{Hawking:1976ra}: states that fall into black hole horizon are entangled with those of Hawking radiation, but unitarity implies that the end state of black hole evaporation (= early+late Hawking radiation) should be a pure state. Authors of \cite{Almheiri:2012rt} argue that (for reasons very close to Mathur's arguments \cite{Mathur:2009hf} , or earlier arguments, e.g., in \cite{Braunstein:2009my,Sorkin:2005qx} and references therein)  the most ``conservative'' resolution is to replace the horizon by a {\it firewall} that ``burns'' infalling observers. Nevertheless, many counter-arguments (and some retractions) soon followed this proposal (e.g. \cite{2012arXiv1207.4090S,2012arXiv1207.5192B,2012arXiv1207.6243H,2012arXiv1207.6626N}).

\begin{figure*}
\centering
\includegraphics[width=\linewidth]{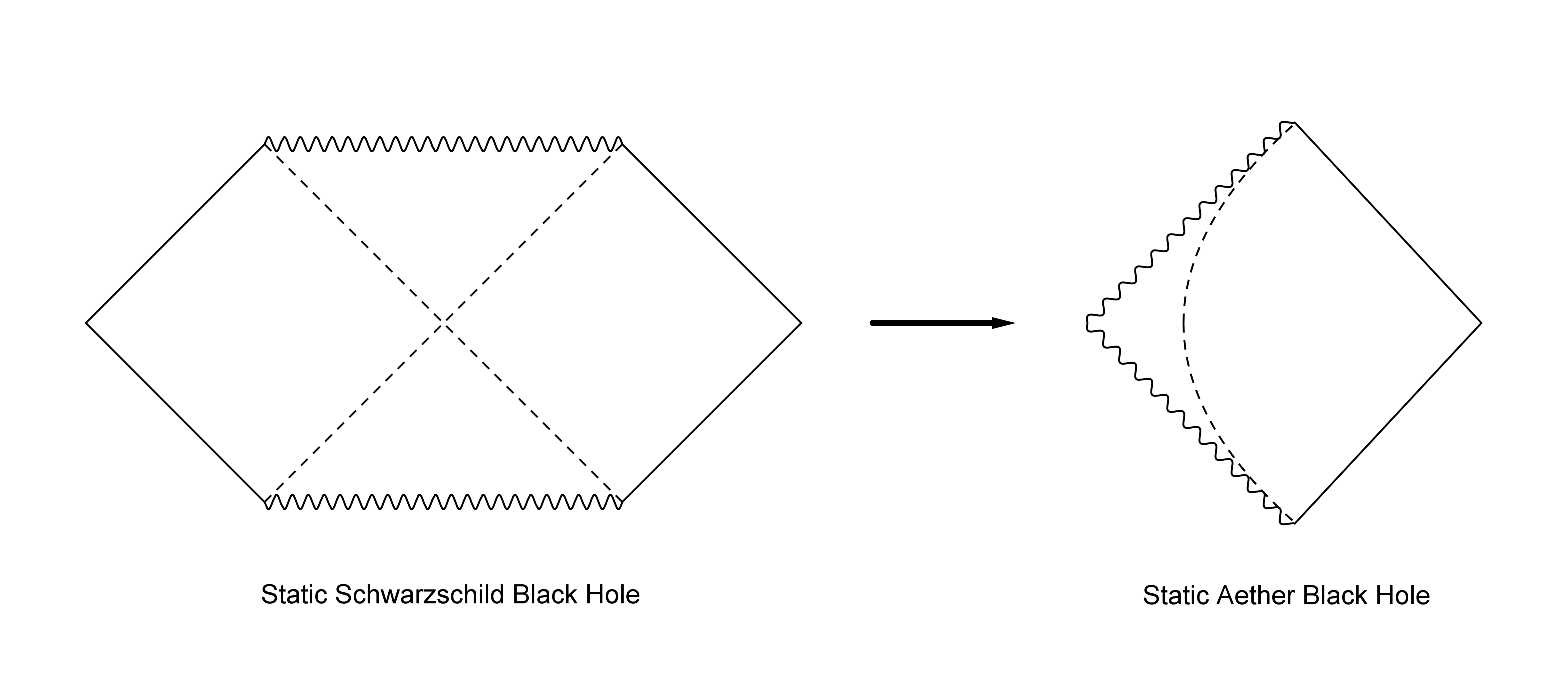}
\caption{Comparison of the causal diagrams for the static Schwarzschild black hole, and the static black holes in gravitational aether   \cite{PrescodWeinstein:2009mp}. In both diagrams, the solid lines depict null infinities, while the squiggly lines are singularities, and $g^{rr}$ vanishes on dotted lines. However, the latter is a null surface in Schwarzschild BH which coincides with event horizon, while it is time-like in the aether BH and corresponds to a throat or minimal area surface.  Moreover, while the singularity is space-like and sits at zero area deep inside the horizon in the Schwarzschild BH, it is null in the Aether BH and sits at finite area, roughly  a Planck length inside the throat. The latter assumption is the key ingredient for aether pressure to match the observed dark energy pressure for astrophysical BH masses  \cite{PrescodWeinstein:2009mp}.}
\label{Static_BH}
\end{figure*}

There is also another argument, rooted in the quantum nature of gravity, for the breakdown of semi-classical spacetime at black hole horizons, which (as shown here) is further validated by the first derivation (to our knowledge) of Bekenstein-Hawking  entropy  based on {\it local} thermodynamics. The argument follows from the behaviour of an {\it incompressible fluid} in GR, which has been argued to develop singularities close to event horizons, while simultaneously explaining the observed scale of cosmological dark energy without any fine-tuning  \cite{PrescodWeinstein:2009mp}. The structure of these black hole solutions is depicted in Fig. (\ref{Static_BH}).

{\it  But why an incompressible fluid?} 
The reason comes from an attempt to solve the (old) cosmological constant problem, which is arguably the most puzzling aspect of coupling gravity to relativistic quantum mechanics \cite{Weinberg:1988cp}.  Given that the natural expectation value for the vacuum of the standard model of particle physics is $\sim 60$ orders of magnitude heavier than the gravitational measurements of vacuum density, it is reasonable to entertain an alternative theory of gravity where the standard model vacuum decouples from gravity. Such a theory could be realized by coupling gravity to the traceless part of the quantum mechanical energy-momentum  tensor. However, the consistency/covariance of gravitational field equations then  requires introducing an auxiliary fluid, the so-called {\it gravitational aether}  \cite{Afshordi:2008xu}. The simplest model for gravitational aether is an incompressible fluid (with vanishing energy density, but non-vanishing pressure), which is currently consistent with all cosmological, astrophysical, and precision tests of gravity \cite{Kamiab:2011am,Aslanbeigi:2011si}: 
\bea
\frac{3}{32 \pi G_N}G_{\mu \nu} =T_{\mu \nu}-\frac{1}{4}T^{\alpha}_{\alpha}g_{\mu \nu}+T'_{\mu \nu},\label{f1}\notag\\
T'_{\mu \nu} =p'(u'_{\mu}u'_{\nu}+g_{\mu \nu}),~
T^{\mu \nu}_{\phantom{\mu \nu};\nu}=0,
\eea
where $G_N$ is Newton's constant, $T_{\mu \nu}$ is the matter energy momentum tensor and $T'_{\mu \nu}$ is the incompressible  gravitational aether fluid. In vacuum, the theory reduces to GR coupled to an incompressible fluid.

Motivated by the existence of singularities or minimum area surfaces close to black hole event horizons in the models mentioned above, we propose a new model for black holes in which spacetime ends, and is orbifolded, at a microscopic distance from black hole horizons. In this model, a black hole is a ``bubble of nothing'' (reminiscent of \cite{1982NuPhB.195..481W}). We then show that putting a  (2+1 dimensional) surface fluid with vanishing density (i.e. incompressible) at the new boundary, which is required by Israel junction conditions, gives a thermodynamic entropy identical to the Bekenstein-Hawking entropy. This work also suggests an analogy between black holes in 3+1 dimensions and 2+1 dimensional incompressible fluids, which have been studied earlier in the context of holography \cite{paul,strominger}.

The outline of our paper is as follows. Section \ref{empty} is devoted to our new model of empty black holes (EBHs) and the resulting local derivation of the Bekenstein-Hawking entropy  for charged rotating empty black holes. We then present a toy microscopic description for the surface fluid with (near-)vanishing density in Section \ref{microscopic}. In section \ref{WhyUnruh}, we comment on the effect of putting a physical boundary at the stretched horizon of a black hole, and discuss how Hawking radiation and Bekenstein-Hawking area law can arise in a realistic setting. Finally, Section \ref{conclude} concludes the paper.

\section{Empty Black Holes and the Bekenstein-Hawking Entropy} \label{empty}
In the previous section, we discussed motivations to posit   a minimum area surface close to/at black hole horizons. As a result, we may model a black hole as a hole in spacetime -- a bubble of nothing --  and end spacetime at a microscopic distance from the putative horizon (at stretched horizon). It will be the responsibility of a full quantum gravity theory to resolve this singularity.

First, we will explain the empty black hole model for spherically symmetric black holes (Schwarzschild) and derive Bekenstein-Hawking entropy. Then, we will extend the model to the most general black holes in four dimensions, i.e. Kerr-Newman black holes.

\subsection{Schwarzschild Black Holes}

Once there is a boundary in spacetime, we need to specify a boundary condition. We impose radial $Z_2$ symmetry at the boundary, as it is a natural boundary condition for a spherically symmetric solution. This  $Z_2$ boundary condition also appears in the membrane paradigm for black holes \cite{1988SciAm.258...69P}.

Consider a static spherically symmetric spacetime which, in general, has the following line element 
\beq\label{f2}
ds^2=-N^2(r)dt^2+\frac{dr^2}{f(r)}+r^2d\Omega^2,
\eeq
where $d\Omega^2\equiv d\theta^2+\sin^2(\theta)d\phi^2$ is the line element of 2-sphere. Here $N(r)$ and $f(r)$ are arbitrary functions  satisfying GR field equations in the bulk (for Schwarzschild $N^2(r)=f(r)=1-\frac{2m}{r}$.) If there were no boundary, then we would have a horizon at $r=r_0$ where $N(r_0)=0$. However, in this model, spacetime ends at a microscopic distance $r=r^*>r_0$  from the horizon.

We assume there is a thin layer of fluid -- not aether -- sitting at the boundary ($r=r^*$), with the following energy-momentum tensor
\beq \label{e1}
\mathcal{T}_{ab}=(\Sigma+\Pi)U_aU_b+\Pi h_{ab},
\eeq
where $\mathcal{T}_{ab}$ is the surface energy-momentum tensor, $\Sigma$ is surface energy density, $\Pi$ is surface pressure, $U_a$ is the fluid 3-velocity, $h_{ab}$ is the induced metric on the hypersurface $r=r^*$ and $a,b\in \{t,\theta,\phi\}$. In fact, as we will show later, imposing radial $Z_2$ symmetry on the boundary {\it requires} the existence of this fluid.


For a general hypersurface $S_r$ defined as $r=\text{constant}>r^*$, the line element on $S_r$ can be written as
\beq
dl_3^2=-N^2(r)dt^2+r^2d\Omega^2.
\eeq
So we obtain 
\beq
h_{ab}=\text{diag}(-N^2,r^2,r^2\sin^2 \theta),
\eeq
and
\beq \label{e2}
K_{ab}=\text{diag}(-N'Nf^{1/2},rf^{1/2},rf^{1/2}\sin^2 \theta),
\eeq
for the extrinsic curvature $K_{ab} = h_a^c h_b ^d \nabla_c n_d$ of the hypersurface, where $N'\equiv \frac{dN}{dr}$.

We now employ the Israel junction conditions
\beq
[h_{ab}] = 0 \label{e0}
\eeq
\beq
[K_{ab}]-[K]h_{ab}=-8\pi \mathcal{T}_{ab}, \label{e3}
\eeq
where $[A]\equiv A(r^+)-A(r^-)$ is the discontinuity of $A(r)$ across the hypersurface $S_r$. Using equations \eqref{e1}, \eqref{e2}, \eqref{e0} and \eqref{e3} for a static solution ($U_a$ having only a non-zero temporal component), we obtain
\bea
[N]&=&0,\\
4\pi \Sigma &=& -[\frac{f^{1/2}}{r}],\\
8\pi \Pi &=& [\frac{N'}{N}f^{1/2}]+[\frac{f^{1/2}}{r}]
\eea
for a hypersurface of radius $r$. In particular, we could write the previous junction conditions at $r=r^*$. However, $S_{r^*}$ is the boundary of spacetime, and the discontinuity of functions across $S_{r^*}$ is not defined. Despite this fact, we can show (see Appendix \ref{app1} for details) that imposing radial $Z_2$ symmetry (for time-like boundaries) modifies \eqref{e0} and \eqref{e3} to 
\beq\label{kab}
K_{ab}-Kh_{ab}=-8 \pi \mathcal{T}_{ab}.
\eeq
As a result, we get
\bea
4\pi \Sigma&=&-\frac{\sqrt{f(r^*)}}{r^*},\label{e5}\\
8\pi \Pi&=&\frac{N'(r^*)}{N(r^*)}\sqrt{f(r^*)}+\frac{\sqrt{f(r^*)}}{r^*}.\label{e6}
\eea
Note that for Schwarzschild metric, in the limit $r^*\rightarrow r_0$ equation \eqref{e5} gives $\Sigma=0$.

Since the surface fluid is at  constant radius it consequently sees the thermal radiation due to its acceleration (Unruh effect \cite{Unruh:1976db}; we will further justify this choice in \ref{conclude}). Assuming the fluid is in thermal equilibrium with the Unruh radiation, its temperature is fixed by the temperature of the radiation in the fluid's vicinity, and so
\beq\label{e7}
T(r^*)=T_{\rm Unruh}=\frac{1}{2 \pi}\frac{N'(r^*)}{N(r^*)}\sqrt{f(r^*)}.
\eeq
Note that the fluid pressure \eqref{e6} and temperature \eqref{e7} diverge in the limit $r^*\rightarrow r_0$.

We now have everything to calculate the entropy of the surface fluid. Assuming local thermodynamic equilibrium (LTE) at zero chemical potential (which is expected at high temperatures), the entropy per unit area of this fluid is given by
\beq\label{e8}
s=\frac{\Sigma+\Pi}{T}
\eeq
yielding
\beq
s=\frac{1}{4},
\eeq
which is the same as Bekenstein-Hawking entropy.

\subsection{Kerr-Newman Black Holes}
We can also extend this model to charged rotating black holes. As shown in Appendix \ref{app2} (see Eq. \ref{ap2}), the near horizon geometry of a Kerr-Newman black hole is 
\beq
ds^2=-\Gamma_+ \lambda^2 d\tau^2+\Gamma_+d\lambda^2+\Gamma_+d\theta^2+\frac{\sin^2\theta}{\Gamma_+}d\psi^2,\notag
\eeq
where $ \Gamma_+ \equiv r_+^2+a^2 \cos^2 \theta$ and 
 $\lambda=0$ corresponds to the horizon of black hole. Similar to the Schwarzschild case, we end the spacetime at the stretched horizon ($\lambda=\lambda^*>0$, taking the limit $\lambda^* \to 0$) and impose $Z_2$ boundary condition at the new boundary.

The $Z_2$ symmetric boundary requires that the extrinsic curvature of the boundary should satisfy \eqref{kab}. Expressing the induced metric on the boundary in $(\tau,\theta,\psi)$ coordinates we obtain
\beq
h_{ab}=\text{diag}(-\Gamma_+ \lambda^{*2},\Gamma_+,\frac{\sin^2 \theta}{\Gamma_+})
\eeq
where the normal vector to the hypersurface $\lambda=\lambda^*$ is
\bea
n_{\mu}=\sqrt{\Gamma_+}(0,1,0,0),\\
n^{\mu}=\frac{1}{\sqrt{\Gamma_+}}(0,1,0,0) .
\eea
Consequently we find
\bea
&K_{ab}=\frac{1}{2\sqrt{\Gamma_+}}(\frac{\partial h_{ab}}{\partial \lambda})_{\lambda^*}=\text{diag}(-\sqrt{\Gamma_+}\lambda^*,0,0),\\
&K= K_{ab}h^{ab}=\frac{1}{\sqrt{\Gamma_+}\lambda^*}.
\eea
Using (\ref{e1}) with
\beq
U_a=\sqrt{\Gamma_+}\lambda^*(1,0,0)
\eeq
(so that $U^a U^b h_{ab} = -1$), we find
\beq
\mathcal{T}_{ab}=\text{diag}(\Sigma\, \Gamma_+ {\lambda^*}^2,\Pi\, \Gamma_+,\Pi \frac{\sin^2 \theta}{\Gamma_+}).
\eeq
Note that this means that the fluid is comoving with the black hole, where  $\sqrt{\Gamma_+} \lambda^*$ is the normalization factor.
Recall that we are working in $(\tau,\theta, \psi)$ coordinates, so a zero velocity component in the direction of $\psi$ means that the fluid is rotating with angular frequency $\Omega$ in the direction of $\phi$.

Using \eqref{kab}, we get
\beq
\Sigma = 0, \qquad
\Pi = \frac{1}{8\pi \sqrt{\Gamma_+}\lambda^*}.
\eeq
As before, we assume equilibrium, and so the temperature of the surface fluid is fixed by the temperature of Unruh radiation.  Since the acceleration of the fluid is
\beq
a=\frac{1}{\sqrt{\Gamma_+} \lambda^*},
\eeq
we find  
\beq
T=a/2 \pi=\frac{1}{2 \pi \sqrt{\Gamma_+}\lambda^*},
\eeq
for~the~Unruh~temperature. Note that this angle-dependent temperature is the same as blue-shifted Hawking temperature of Kerr-Newman black hole for a co-rotating observer at the position of the boundary.
Finally, the entropy per unit area of the fluid will be 
\beq
s=\frac{\Sigma+\Pi}{T}=\frac{1}{4}.
\eeq


\section{Microscopic Derivation of an Incompressible Fluid}\label{microscopic}
In the previous section, we showed that a stationary solution requires the existence of an incompressible surface fluid on the boundary of an EBH. In this section we  show that a class of dispersion relations for matter can give rise to a nearly incompressible fluid at high energies (by nearly incompressible, we mean that the pressure of the fluid is much greater than its energy density).

For a thermal gas of bosons/fermions at temperature $T$, energy density $\rho$ and pressure $\mathcal{P}$ are as follows
\bea
\rho(T) &=&g \int \frac{d^3p}{(2\pi)^3}E(p)n(p)=<E>,\label{t1}\\
\mathcal{P}(T)&=&\frac{1}{3}g\int \frac{d^3p}{(2\pi)^3}p\frac{dE}{dp}n(p)=\frac{1}{3}<p\frac{dE}{dp}>,\label{t2}
\eea
where $g$ is the degeneracy factor, $E(p)$ is dispersion relation (relation between energy and momentum of a particle) and
\beq
n(p)=\frac{1}{e^{E(p)/T}\pm 1},
\eeq
where for simplicity we set the chemical potential $\mu=0$ ($+$ for fermions, $-$ for bosons.)

Once we specify a dispersion relation, we are able to compute the energy density and pressure of a fluid of these particles. While particles with mass $m$ at low energies satisfy the Lorentzian dispersion relation $E^2=p^2+m^2$, as we argue below, there are reasons to believe the energy-momentum relation might be modified at high energies (e.g. \cite{Horava:2009uw}).

\begin{figure}[h!]
\centering
\includegraphics[width=\linewidth]{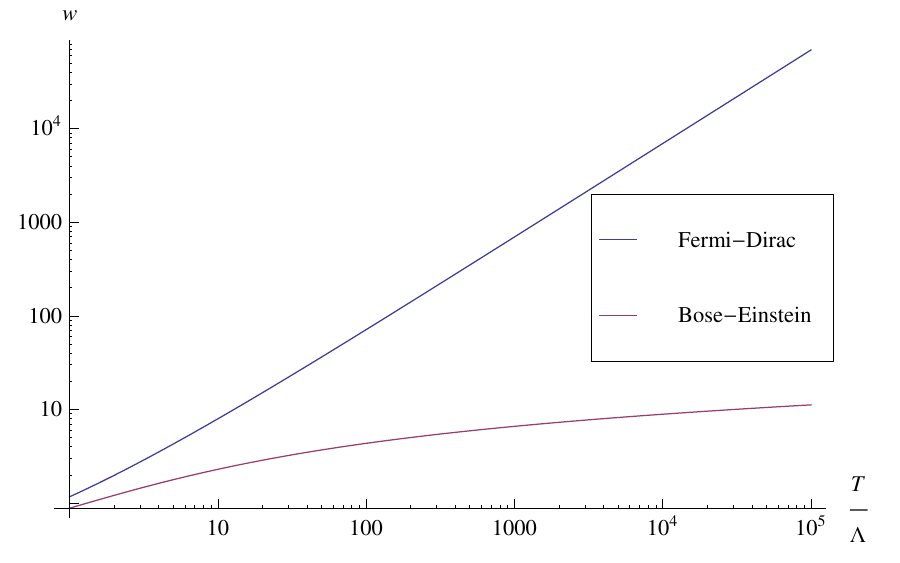}
\caption{Equation of State variable $(w=\frac{\mathcal{P}}{\rho})$ as a function of Temperature.}
\label{fig1}
\end{figure}

For example, consider the dispersion relation \cite{Lee}
\beq\label{t3}
E^2=\frac{p^2}{1-p^2/\Lambda^2},
\eeq
which reduces to the Lorentzian dispersion relation (with $m=0$) at low energies ($p \ll \Lambda$), while it deviates from Lorentzian dispersion relation at high energies. Indeed, energy becomes infinite for a finite value of momentum. In Fig.\ref{fig1} we depict the equation of state variable $w$ (pressure over density) of a fluid obeying the dispersion relation \eqref{t3}, as a function of temperature.
We see that $w$ grows as temperature increases, and approaches infinity in the limit of infinite compression, expected for the surface fluid. 

Generically, fluid of particles with a dispersion relation in which energy as a function of momentum grows faster than any power law becomes a near incompressible fluid at high temperatures. We will show this by an example. Consider the following dispersion relation
\beq
E^2=p^2+\frac{p^{2l}}{\Lambda^{2(l-1)}}.
\eeq
At low temperatures $(T \ll \Lambda)$, energies of particles are effectively equal to their momentum $E \approx p$, and according to \eqref{t1} and \eqref{t2}
\beq
w=\frac{\mathcal{P}}{\rho} \approx \frac{1}{3}.
\eeq
However, at high temperatures $(T \gg \Lambda)$ the dispersion relation changes to  $E \approx \frac{p^l}{\Lambda^{(l-1)}}$ and, as a result
\beq\label{t5}
w \approx \frac{l}{3}.
\eeq
However, if dispersion relation grows faster than any power law for large $p$, then $l$ increases as temperature increases and goes to infinity as temperature goes to infinity. Consequently, according to \eqref {t5}, $w$ increases unboundedly with temperature.

On the other hand, the dispersion relation \eqref{t3} regulates the UV infinities of quantum field theory, since there is a maximum momentum for any particle. We show here the argument for a real scalar field $\phi$. In the interaction picture, $\phi$ can be expanded in terms of creation and annihilation operators as 
\beq
\phi(x)=\int^{\Lambda} \frac{d^3\vec p}{(2\pi)^3}\frac{1}{\sqrt{2E_{\vec p}}}(a_{\vec p}e^{ip \cdot x}+a^{\dag}_{\vec p}e^{-ip \cdot x}),
\eeq 
where $p\cdot x \equiv -p^0t+\vec p \cdot \vec x$, $p^0=E_{\vec p}( \equiv E(p))$ and $[a_{\vec p},a^{\dag}_{\vec q}]=(2 \pi)^3 \delta^{(3)}(\vec p -\vec q)$. Using the above expansion, we obtain 
\bea
G_F(x-y)=\langle 0|T \phi(x)\phi(y)|0\rangle \nonumber\\ = i\int \frac{d^4p}{(2\pi)^4}\frac{\theta(\Lambda-|\vec p|)}{(p^0)^2-(E_{\vec p})^2+i \epsilon}e^{ip\cdot (x-y)}.
\eea

For canonical interactions, loop infinities originate from integration over products of Feynman Green's functions. However, since there is a cut off for spatial part of momentum, loop corrections of quantum field theory will be finite.

\section{Why Unruh Temperature?}\label{WhyUnruh}

 Our derivation of EBH entropy crucially relied on assuming that the surface fluid is heated up to the Unruh temperature \cite{Unruh:1976db}, set by the acceleration of the observers on the stretched horizon. The original derivation of Unruh (Hawking) radiation relies on a (locally) Minkowski vacuum at the vicinity of the horizon, and altering the dispersion relation (Sec. \ref{microscopic}) at high momenta will change the properties of the vacuum state. As a result,  on small scales one no longer expects the Minkowski vacuum to be an adequate description. However, further investigation \cite{Unruh:1994zw, Unruh:2004zk} showed that generally the Hawking radiation does not heavily depend on the dispersion relation at high momenta (see \cite{Unruh:2004zk} for conditions and exceptions), and the spectrum at energies below the UV scale, $\Lambda$ (e.g., Planck energy) remains unchanged to lowest order in powers of $E/\Lambda$. 

Similarly, one might suspect that the existence of a physical boundary at the stretched horizon might change the radiation process. Here, we demonstrate that putting a boundary at the stretched horizon will not change Hawking radiation to the lowest order.  Reduced to its bare bones,  the Unruh temperature is simply a statement of Heisenberg uncertainty principle: the temperature of an emitting region should be bigger than the inverse of its size. Any lower temperature cannot be in local thermal equilibrium (LTE), which was our key assumption preceding Eq. (\ref{e8}). In order to show this rigorously, we first briefly review a simple derivation of Hawking radiation similar to \cite{NouriZonoz:1998td}.

A massless scalar field $\Phi$ in Schwarzschild spacetime satisfies the following wave equation
\beq\label{t1}
\frac{1}{\sqrt{-g}}\partial_{\mu}\left(\sqrt{-g}~g^{\mu \nu} \partial_{\nu}\Phi\right)=0.
\eeq
Using Kruskal coordinates
\bea
v&=&\left|\frac{r}{2m}-1\right|^{\frac{1}{2}}e^{\frac{r}{4m}}\notag\\
&\times&\left[\sinh \left(\frac{t}{4m}\right) \theta(r-2m)
+\cosh\left(\frac{t}{4m}\right) \theta(2m-r)\right],\notag\\
u&=&\left|\frac{r}{2m}-1\right|^{\frac{1}{2}}e^{\frac{r}{4m}}\notag\\
&\times&\left[\cosh \left(\frac{t}{4m}\right) \theta(r-2m)+\sinh\left(\frac{t}{4m}\right) \theta(2m-r)\right],\notag
\eea
where $\theta(x)$ is the Heaviside step function, the metric becomes 
\beq
ds^2=\frac{32 m^3}{r}e^{-\frac{r}{2m}}(-dv^2+du^2)+r^2d\Omega^2.
\eeq
Considering only the radial mode $\partial_{\theta} \Phi=\partial_{\varphi} \Phi=0$, Equation \eqref{t1} results in 
\beq\label{t2}
-\partial_v \left(r^2\partial_v \Phi\right)+\partial_u \left(r^2\partial_u \Phi \right)=0,
\eeq
Where $r$ is an implicit function of $u$ and $v$. Substituting $\Phi=\frac{1}{r}\Psi$ into \eqref{t2}, we get
\beq\label{t3}
\left(-\partial_v^2 +\partial_u^2 -\frac{64m^4}{r^4}e^{-\frac{r}{2m}}\right)\Psi=0.
\eeq
Plane-waves $e^{ik(u-v)}$ are outgoing solutions  to the above equation provided that $k^2 \gg \frac{64m^4}{r^4}e^{-\frac{r}{2m}}$. Since we are considering the outgoing modes outside horizon, the term $\frac{64m^4}{r^4}e^{-\frac{r}{2m}}$ reaches its maximum at $r=2m$, which is of order unity. As a result, for $k \gg 1$, i.e. wavelengths much shorter than the size of the black hole, $\Phi_k=\frac{1}{r}e^{ik(u-v)}$ are valid solutions outside the horizon.  Note that the non-radial modes satisfy a similar equation to \eqref{t3} with a different potential term. For the case of non-radial modes or $k \lesssim 1$, the potential term only introduces a gray body factor in the last result \cite{Mukhanov:2007zz}, which is of no interest in the present calculation.

A static observer in Schwarzschild coordinates $(r,t)$ decomposes this wave into its Fourier modes $\tilde \Phi_k(\omega)$ using time coordinate $t$ (which is its proper time, up to a gravitational redshift factor). As a result, 
\beq
\tilde \Phi_k(\omega)=\int_{-\infty}^{+\infty} dt ~e^{-i \omega t}\Phi_k(r,t)=\frac{1}{r}\int_{-\infty}^{+\infty} dt~ e^{-i \omega t}e^{i \sigma e^{-\frac{t}{4m}}},
\eeq
where we have used $u-v=(\frac{r}{2m}-1)^{\frac{1}{2}}e^{\frac{r}{4m}}e^{-\frac{t}{4m}}$ and $\sigma \equiv k(\frac{r}{2m}-1)^{\frac{1}{2}}e^{\frac{r}{4m}}>0$. After a straightforward calculation, we get
\beq\label{t7}
\tilde \Phi_k(\omega)=\frac{4m}{r} (-i\sigma)^{-4im\omega}\Gamma(4im\omega),
\eeq
which results in the following power spectrum
\beq\label{t4}
|\tilde \Phi_k(\omega)|^2=\frac{8 \pi m}{r^2\omega}\frac{1}{e^{8 \pi m \omega}-1}.
\eeq
This shows that, up to gray body factors and gravitational redshift, the observer in Schwarzschild coordinates detects   thermal radiation with temperature $T=T_{H}=\frac{1}{8 \pi m}$.

Now, let's assume we cut off the solution $\Phi_k=\frac{1}{r}e^{ik(u-v)}$ at some minimum radius $r_{min}=2m(1+\epsilon)$ on a $v=0$ hypersurface (for simplicity), and replace it with 
\beq\label{t5}
\Phi^c_{0,k}(u,v=0)=\frac{1}{r}e^{i k u} \theta(r-r_{min}).
\eeq
Using 
\beq
u^2-v^2=\left(\frac{r}{2m}-1\right)e^{\frac{r}{2m}},
\eeq
on $v=0$, we get
\beq
\theta(r-r_{min})=\theta(u-u_{min}),
\eeq
where $u_{min} \equiv (\frac{r_{min}}{2m}-1)^{\frac{1}{2}}e^{\frac{r_{min}}{4m}}$.
As a result, the outgoing solutions to \eqref{t3} with initial conditions \eqref{t5} will be
\beq
\Phi^c_k(u,v)=\frac{1}{r}\Psi^c_{0,k}(u-v,v=0)=\frac{1}{r}e^{ik(u-v)}\theta(u-v-u_{min}).
\eeq
Expressing $\Phi^c_k$ in terms of Fourier modes of the Schwarzschild observer, we get
\bea
\tilde \Phi^c_k(\omega)&=&\int_{-\infty}^{+\infty} dt ~e^{-i \omega t}\Phi^c_k(r,t)\notag\\
&=&\frac{1}{r}\int_{-\infty}^{+\infty} dt ~e^{-i \omega t}e^{i \sigma e^{-\frac{t}{4m}}}\theta\left[\left(\frac{r}{2m}-1\right)^{\frac{1}{2}}e^{\frac{r-t}{4m}}-u_{min}\right]\notag\\
&=&\frac{1}{r}\int_{-\infty}^{+\infty} dt ~e^{-i \omega t}e^{i \sigma e^{-\frac{t}{4m}}}\theta(t_{max}-t),\label{t6}
\eea
where $e^{-\frac{t_{max}}{4m}}=u_{min}\left(\frac{r}{2m}-1\right)^{-\frac{1}{2}} e^{-\frac{r}{4m}}=\frac{\left(\frac{r_{min}}{2m}-1\right)^{\frac{1}{2}} e^{\frac{r_{min}}{4m}}}{\left(\frac{r}{2m}-1\right)^{\frac{1}{2}} e^{\frac{r}{4m}}}$.

Using
\beq
\theta(t)=-\frac{1}{2 \pi i}\int_{-\infty}^{+\infty}\frac{df}{f+i a}e^{-i f t},
\eeq
where $a$ is a small positive number, together with \eqref{t7}, we get
\beq\label{t8}
\begin{split}
\tilde \Phi^c_k(\omega)=\int_{-\infty}^{+\infty} -\frac{df}{2 \pi i } \left(\frac{1}{f+i a}\right)e^{-i f t_{max}}\left(\frac{4m}{r}\right)\\
\times(-i\sigma)^{4im(f-\omega)}\Gamma\left[4im(\omega-f)\right].
\end{split}
\eeq
Closing the contour in the lower half complex plane of $f$ (note that $t_{max}>0$), we can express \eqref{t8} in terms of residues of the integrand. Contribution from the pole at $f=-ia$ is
\beq
\frac{4m}{r}(-i\sigma)^{-4im\omega}\Gamma(4im\omega)=\tilde \Phi_k(\omega).
\eeq
The Gamma function $\Gamma(z)$ has also simple poles at $z=-n$ where $n=0,1,...$ with residues $\frac{(-1)^n}{n!}$; their contribution is
\beq
\tilde \Phi^{(n)}(\omega)\equiv-\frac{4m}{r}\frac{e^{- i \omega t_{max}}}{n+4i m \omega}\frac{\left(i \sigma e^{-\frac{t_{max}}{4m}}\right)^n}{n!}.
\eeq
 Despite the fact that the pole $f=\omega$ corresponding to $n=0$ is on the real line and must be treated more carefully,
 we argue that either there is no contribution from this pole or that this contribution is not important. First, the contribution from the $n=0$ pole is the same as adding a constant to $\Phi^c_k(r,t)$. Since the field $\Phi$ is massless, adding a constant term has no observational effect and can be ignored. On the other hand, we expect to recover the result of the previous calculation for $\tilde \Phi_k(\omega)$ in the limit $t_{max} \rightarrow +\infty$. This fact also shows that the $n=0$ pole should not contribute to $\tilde \Phi^c_k(\omega)$. Finally, we get
\bea
&~&\tilde \Phi^c_k(\omega)=\tilde \Phi_k(\omega)+\sum_{n=1}^{\infty}\tilde\Phi^{(n)}(\omega)\notag\\
&=&\tilde \Phi_k(\omega)-\frac{4m}{r}e^{- i \omega t_{max}}\sum_{n=1}^{\infty}\frac{\left(i \sigma e^{-\frac{t_{max}}{4m}}\right)^n}{(n+4i m \omega)n!}.
\eea
The new contributions are suppressed with the power of $\sigma e^{-\frac{t_{max}}{4m}}=k\left(\frac{r_{min}}{2m}-1\right)^{\frac{1}{2}}e^{\frac{r_{min}}{4m}} \approx k (\epsilon e)^{\frac{1}{2}}$ for small value of $\epsilon$. 

For $k\epsilon^{\frac{1}{2}} \ll 1$, the leading contribution from the corrections comes from $n=1$ term. Comparing this term with the thermal part $\tilde \Phi_k(\omega)$, we get
\beq\label{phi1}
\left|\frac{\tilde\Phi^{(1)}(\omega)}{\tilde \Phi_k(\omega)}\right|^2=\frac{e}{2\pi}k^2\epsilon \frac{4m\omega}{1+16m^2\omega^2}\left(e^{8\pi m \omega}-1\right).
\eeq
This shows that the corrections are not important for $\omega/T_{H} \lesssim -\ln(k^2 \epsilon)$. However, because of exponentially damping term in thermal power spectrum, the corrections become important for higher frequencies. Consequently, putting a physical boundary at the stretched horizon, only changes the tail of the thermal power spectrum.  

 Let us now recap what we have done so far: We have constructed a model for a general black hole in Section \ref{empty} which realizes the idea of a firewall on its ``horizon'', and results in a local derivation for Bekenstein-Hawking entropy. The latter is based on the assumption of an Unruh temperature for the firewall, which we justified earlier in this section. However, this model is idealized, as there is a perfect incompressible {\it surface} fluid on the stretched horizon of EBH. In Section \ref{microscopic}, we have proposed a microscopic description for incompressible fluid. Here, we will put these ideas together to construct a more physical picture.
\begin{figure}
\centering
\includegraphics[width=0.75\linewidth]{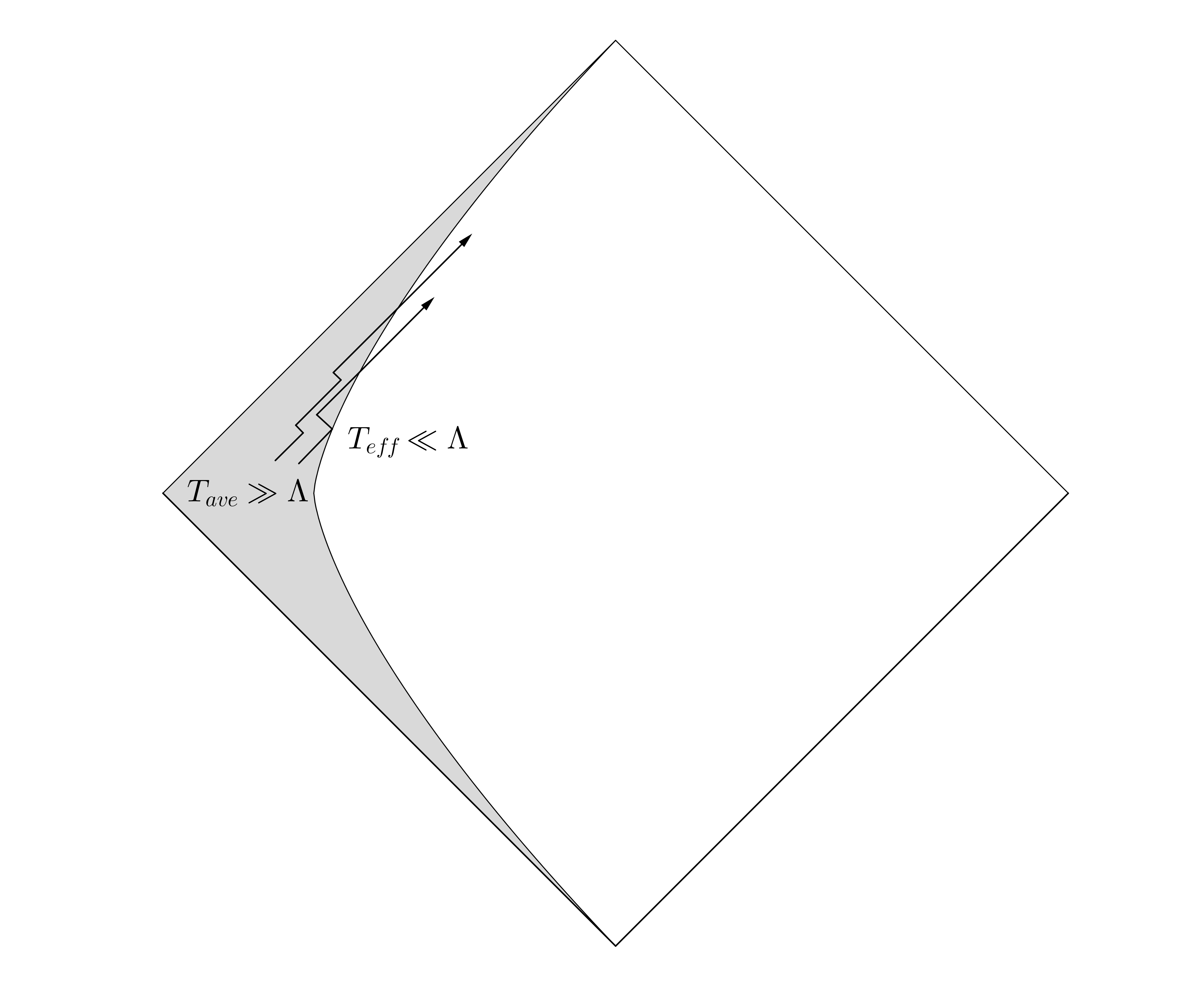}
\caption{ ``Photosphere'' of a Firewall: This figure demonstrates that even though the average temperature of a firewall could be greater than $\Lambda$, where the fluid becomes incompressible, the ``photosphere'' might have an effective temperature less than the Lorentz violation scale $\Lambda$. As a result the derivations of Hawking/Unruh temperature goes through. As we argue in the text, this implies that Bekenstein-Hawking area law is an upper limit for the entropy of the firewall.  }
\label{EBH_Photo}
\end{figure}

 The surface fluid can be thought of  as a thin shell of matter with finite (non-zero) thickness concentrated on the stretched horizon of the EBH (Fig. \ref{EBH_Photo}). Upon accretion, matter is heated up to the firewall average temperature $T_{\rm ave}$. For an ``optically'' thick object, this temperature is {\it higher} than the effective temperature of the ``photosphere'', $T_{\rm eff}$ which is where ``radiation'' leaves the surface of the fluid \footnote{ This is dictated by radiative transfer equations, which is exactly the same reason the average temperature of the sun ($\sim 1.5 \times 10^7$K) is higher than that of its photosphere ($\sim 6 \times 10^3$ K).}. We should note that  ``photosphere'' here refers to the radius where ``optical'' thickness to infinity equals unity, for the particles that are mostly responsible for energy transfer. For example, for a supernova explosion, this is mainly due to neutrinos, as photons become optically thin much farther out, and thus have much lower effective temperatures.  

 We have already shown above that vacuum state defined by Kruskal modes is seen by a Schwarzschild observer as a thermal state with blue-shifted Hawking (or Unruh) temperature. This argument relies on the fact that Lorentz symmetry is still preserved. However, we have changed the dispersion relation to acquire an incompressible fluid on the stretched horizon. The only consistent way to satisfy all these requirements is to assume that $T_{\rm eff}$ is smaller than Lorentz violation energy scale, $\Lambda$. With this assumption, the relativistic derivation above for Hawking/Unruh radiation is still valid.
 
 On the other hand, $T_{\rm ave}$ must be higher than $\Lambda$ in order to get an incompressible fluid on the stretched horizon. Consequently, we require
\beq
T_{\rm Unruh} \simeq T_{\rm eff} \ll \Lambda \ll T_{\rm ave}.
\eeq
This condition is consistent with what we expect from radiative transfer within the firewall, i.e. $T_{\rm ave} > T_{\rm eff}$ to get the outward flux of radiation.

Consider next the entropy equation,
\beq
S=\frac{\Pi}{T}.
\eeq
Previously, we derived that $S=S_{BH}$ if $T= T_{\rm ave}$ was blue-shifted Hawking (or Unruh) temperature. However, note that $T_{\rm eff}$ is the blue-shifted Hawking temperature and since $T_{\rm ave} > T_{\rm eff}$, we have
\beq
S<S_{BH}.
\eeq
This shows that Bekenstein-Hawking entropy is an {\it absolute upper limit} for the entropy of matter condensation on the firewall. The ratio $\frac{S}{S_{BH}}=\frac{T_{\rm eff}}{T_{\rm ave}}$ depends on how sharply the Lorentz violation {\it and} rise in optical depth happen. For example, in an extreme case one can imagine a very sharp violation of the Lorentz dispersion relation and rise in optical depth, where $T_{\rm ave}$ and $T_{\rm eff}$ can be made very close to $\Lambda$. In this case $\frac{S}{S_{BH}} \rightarrow 1$.

 Finally, we should also mention that $\Lambda$ is the energy scale of Lorentz violation in the matter sector and it can be smaller than Planck energy. This implies that although the fluid is heated up to an internal temperature above $\Lambda$, its temperature can still be smaller than Planck energy. As a result, the classical general relativity description of the Israel junction conditions may still be valid.

\section{Discussion and Conclusions} \label{conclude}
\begin{figure*}
\centering
\includegraphics[width=\linewidth]{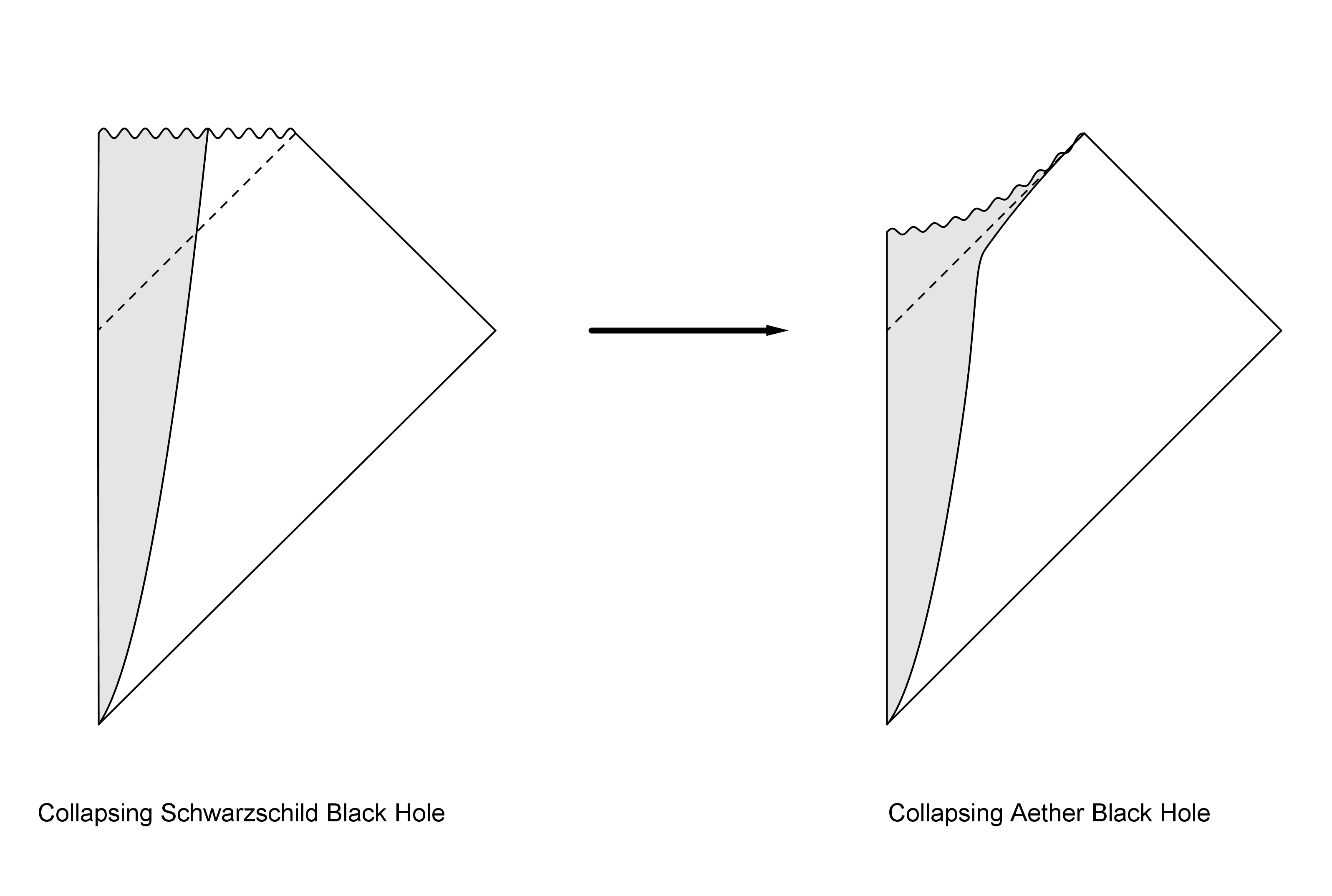}
\caption{Comparison of the causal diagrams for a collapsing Schwarzschild black hole, and our proposed picture for collapsing black holes in gravitational aether or firewall scenarios. In both diagrams, the dotted lines depict classical event horizons, while the squiggly lines are singularities. The black area correspond to the collapsing star. While  in the Schwarzschild BH, the singularity is space-like and deep inside the horizon, in the aether/firewall case it approaches the horizon and becomes null asymptotically. The accreted material smoothly crosses the Schwarzschild horizon,  but it condenses into Planckian densities just inside the horizon of the aether BH.}
\label{Collapsing_BH}
\end{figure*}

We have shown that a surface fluid at Unruh temperature with vanishing energy density (but non-vanishing pressure) on the stretched horizon of a black hole 
-- which we call an empty black hole --
has the same thermodynamic entropy as Bekenstein-Hawking entropy. The surface fluid is the result of ending (or orbifolding) spacetime at the stretched horizon and replacing it with a $Z_2$ symmetric boundary. We therefore conjecture that the microstates of a black hole are those of the surface fluid at the stretched horizon. We emphasize that this description is very similar to the traditional membrane paradigm for black hole horizons \cite{1988SciAm.258...69P}. However, in our description, the membrane properties are physical and result from condensation of accreted matter onto a {\it physical} membrane, while they arise only as a mathematical analogy in the traditional membrane paradigm (e.g. the pressure in \cite{Kolekar:2011gw}). Although the EBH model was constructed in vacuum, there is a singularity (or boundary) at the horizon. This situation could  arise via  some unknown quantum gravity effects such as a firewall or via gravitational aether. The latter has the advantage that it is a classical theory and so in principle an analysis of the collapse of matter in this theory can be carried out. Fig. (\ref{Collapsing_BH}) compares the expected causal diagrams for collapse of standard black holes and  EBH's. 

 Clearly a surface fluid with vanishing energy, but non-vanishing pressure (i.e. incompressible) necessitates invoking an exotic phase of matter. Interestingly, we notice that a field theory with a Lorentz violating momentum cut-off, $\Lambda$, which regulates all UV divergences of canonical quantum field theories, approaches this equation of state at high temperatures. In other words, an (admittedly na\"ive) UV completion of the quantum field theory, also reproduces the correct entropy of the black hole. Finally, in the last section we showed that, even if Lorentz symmetry is violated in the firewall, to ensure its incompressibility, its photosphere could still be much cooler than the scale $\Lambda$, and thus emit the canonical Hawking spectra. This sets the Bekenstein-Hawking area law as a strict upper limit for the entropy of the firewall.

To our knowledge, our derivation of Bekenstein-Hawking entropy is the first truly {\it local} description of black hole micro-states.  For example, models for the area law based on entanglement entropy, or string theory (e.g., in the fuzzball proposal) are inherently non-local: the total entanglement entropy depends on an uncertain cut-off, and only reproduces the area law up to an order unity factor \cite{Bombelli:1986rw}. Computing the change in the entanglement entropy can reproduce the correct change in the area law (assuming Einstein equations) \cite{Jacobson:1995ab, Bianchi:2012br}, but does NOT localize the total entropy on the surface. Similarly, the counting of string theory fuzzball solutions involves summing over states with no classical 4d counterpart (e.g., \cite{Mathur:2008kg}). 

Another proposal for the endpoint of gravitational collapse, which is similar to our model, is the gravastar \cite{gravastar}. The exterior of a gravastar is the Schwarzschild geometry, whereas the interior is replaced by a de Sitter spacetime that is matched to the exterior via a thin shell of stiff fluid (a fluid with $p=\rho$.) However, the entropy of a gravastar is much less than the Bekenstein-Hawking entropy.

 Additionally, the Bekenstein-Hawking entropy can be derived as the thermodynamic entropy of a stiff fluid in the presence of a black hole
\cite{zurek, t'Hooft}. In this model, the horizon is replaced by a high density thin shell of stiff fluid. However, this solution suffers from the presence of a point-like naked singularity with negative mass at the center. In this context, an interesting question for future study would be to consider the gravitational collapse of matter (e.g. a ball of dust) in the more general context of gravitational aether to see how the end state is (classically) approached. 

While the list of motivations for truncating the classical spacetime at black horizons has grown manyfold over the past few years, here we note the recurrence of the notion of {\it incompressibility}, in our exposition. Indeed, incompressibility is no longer considered a mathematical novelty (e.g. \cite{Afshordi:2006ad, Afshordi:2009tt, Bredberg:2010ky, strominger, paul}), and has appeared, and re-appeared in different physical contexts.   In particular, the appearance of an incompressible fluid in both bulk {\it and} boundary of gravitational aether black holes 
is suggestive of the continuity of the underlying microscopic phenomenon. The obvious difference between the two fluids, however, is that the boundary fluid carries a finite entropy, while the gravitational aether has a degenerate phase space (and thus zero or negligible entropy). Nevertheless, it is not clear whether this difference might be an artifact of the surface condensation, or rather point to different origins of the two fluids. A related puzzle is why Bekenstein-Hawking entropy is localized on the surface (at least at the classical level),  while the bulk incompressible fluid can cause non-local interactions.

Furthermore, one might ask what type of horizons are expected to develop singularities. Do we expect singularities for Rindler or de Sitter horizons? For example it has recently been argued \cite{Almheiri:2012rt} that singularities (or firewalls) only occur in ``old'' horizons, at a fraction of their evaporation time, which never happens for de Sitter or Rindler horizons.

Even more speculative, but most exciting, is the possibility of directly probing quantum gravitational effects by precision studies of astrophysical horizons (e.g. \cite{Broderick:2007ek}). Given that the conditions of big bang is now replicated close to black hole horizons (and not deep inside them), the observational constraints on the early universe might now be applied to the microscopic structure of horizons, or conversely, black hole observations can potentially constrain the nature of  cosmological big bang.  

\vspace*{10mm}

{\it Acknowledgement:} The authors would like to thank Siavash Aslanbeigi, Eugenio Bianchi, Samir Mathur, Paul McFadden, Rafael Sorkin, and Dejan Stojkovic for invaluable discussions. This work was supported by the Natural Science and Engineering Research Council of Canada,
the University of Waterloo and the Perimeter Institute for Theoretical Physics. Research
at the Perimeter Institute is supported by the Government of Canada through Industry
Canada and by the Province of Ontario through the Ministry of Research \& Innovation. 

\bibliographystyle{ieeetr}
\bibliography{Empty_BH-v12a}

\begin{thebibliography}{10}

\bibitem{Mathur:2008kg}
S.~D. Mathur, ``{Tunneling into fuzzball states},'' {\em Gen.Rel.Grav.},
  vol.~42, pp.~113--118, 2010.

\bibitem{Mathur:2009zs}
S.~D. Mathur, ``{How fast can a black hole release its information?},'' {\em
  Int.J.Mod.Phys.}, vol.~D18, pp.~2215--2219, 2009.

\bibitem{Mathur:2012zp}
S.~D. Mathur, ``{Black Holes and Beyond},'' 2012.

\bibitem{Hawking:1976ra}
S.~Hawking, ``{Breakdown of Predictability in Gravitational Collapse},'' {\em
  Phys.Rev.}, vol.~D14, pp.~2460--2473, 1976.

\bibitem{Almheiri:2012rt}
A.~Almheiri, D.~Marolf, J.~Polchinski, and J.~Sully, ``{Black Holes:
  Complementarity or Firewalls?},'' 2012.

\bibitem{Mathur:2009hf}
S.~D. Mathur, ``{The Information paradox: A Pedagogical introduction},'' {\em
  Class.Quant.Grav.}, vol.~26, p.~224001, 2009.

\bibitem{Braunstein:2009my}
S.~L. Braunstein, S.~Pirandola, and K.~Zyczkowski, ``{Entangled black holes as
  ciphers of hidden information},'' {\em Physical Review Letters 110,},
  vol.~101301, 2013.

\bibitem{Sorkin:2005qx}
R.~D. Sorkin, ``{Ten theses on black hole entropy},'' {\em
  Stud.Hist.Philos.Mod.Phys.}, vol.~36, pp.~291--301, 2005.

\bibitem{2012arXiv1207.4090S}
L.~{Susskind}, ``{Complementarity And Firewalls},'' 2012.

\bibitem{2012arXiv1207.5192B}
R.~{Bousso}, ``{Observer Complementarity Upholds the Equivalence Principle},''
  July 2012.

\bibitem{2012arXiv1207.6243H}
D.~{Harlow}, ``{Complementarity, not Firewalls},'' July 2012.

\bibitem{2012arXiv1207.6626N}
Y.~{Nomura}, J.~{Varela}, and S.~J. {Weinberg}, ``{Complementarity Endures: No
  Firewall for an Infalling Observer},'' July 2012.

\bibitem{PrescodWeinstein:2009mp}
C.~Prescod-Weinstein, N.~Afshordi, M.~L. Balogh, N.~Afshordi, and M.~L. Balogh,
  ``{Stellar Black Holes and the Origin of Cosmic Acceleration},'' {\em
  Phys.Rev.}, vol.~D80, p.~043513, 2009.

\bibitem{Weinberg:1988cp}
S.~Weinberg, ``{The Cosmological Constant Problem},'' {\em Rev.Mod.Phys.},
  vol.~61, pp.~1--23, 1989.

\bibitem{Afshordi:2008xu}
N.~Afshordi, ``{Gravitational Aether and the thermodynamic solution to the
  cosmological constant problem},'' 2008.

\bibitem{Kamiab:2011am}
F.~Kamiab and N.~Afshordi, ``{Neutron Stars and the Cosmological Constant
  Problem},'' {\em Phys.Rev.}, vol.~D84, p.~063011, 2011.

\bibitem{Aslanbeigi:2011si}
S.~Aslanbeigi, G.~Robbers, B.~Z. Foster, K.~Kohri, and N.~Afshordi,
  ``{Phenomenology of Gravitational Aether as a solution to the Old
  Cosmological Constant Problem},'' {\em Phys.Rev.}, vol.~D84, p.~103522, 2011.

\bibitem{1982NuPhB.195..481W}
E.~{Witten}, ``{Instability of the Kaluza-Klein vacuum},'' {\em Nuclear Physics
  B}, vol.~195, pp.~481--492, Feb. 1982.

\bibitem{paul}
G.~Compere, P.~McFadden, K.~Skenderis, and M.~Taylor, ``{The relativistic fluid
  dual to vacuum Einstein gravity},'' {\em JHEP}, vol.~03, p.~076, 2012.

\bibitem{strominger}
I.~Bredberg and A.~Strominger, ``{Black Holes as Incompressible Fluids on the
  Sphere},'' {\em JHEP}, vol.~1205, p.~043, 2012.

\bibitem{1988SciAm.258...69P}
R.~H. {Price} and K.~S. {Thorne}, ``{The membrane paradigm for black holes},''
  {\em Scientific American}, vol.~258, pp.~69--77, Apr. 1988.

\bibitem{Unruh:1976db}
W.~Unruh, ``{Notes on black hole evaporation},'' {\em Phys.Rev.}, vol.~D14,
  p.~870, 1976.

\bibitem{Horava:2009uw}
P.~Horava, ``{Quantum Gravity at a Lifshitz Point},'' {\em Phys.Rev.},
  vol.~D79, p.~084008, 2009.

\bibitem{Lee}
J.~Magueijo and L.~Smolin, ``{Generalized Lorentz invariance with an invariant
  energy scale},'' {\em Phys.Rev.}, vol.~D67, p.~044017, 2003.

\bibitem{Unruh:1994zw}
W.~Unruh, ``{Dumb holes and the effects of high frequencies on black hole
  evaporation},'' 1994.

\bibitem{Unruh:2004zk}
W.~G. Unruh and R.~Schutzhold, ``{On the universality of the Hawking effect},''
  {\em Phys.Rev.}, vol.~D71, p.~024028, 2005.

\bibitem{NouriZonoz:1998td}
M.~Nouri-Zonoz and T.~Padmanabhan, ``{The Classical essence of black hole
  radiation},'' 1998.

\bibitem{Mukhanov:2007zz}
V.~Mukhanov and S.~Winitzki, ``{Introduction to quantum effects in gravity},''
  2007.

\bibitem{Kolekar:2011gw}
S.~Kolekar and T.~Padmanabhan, ``{Action Principle for the Fluid-Gravity
  Correspondence and Emergent Gravity},'' {\em Phys.Rev.}, vol.~D85, p.~024004,
  2012.

\bibitem{Bombelli:1986rw}
L.~Bombelli, R.~K. Koul, J.~Lee, and R.~D. Sorkin, ``{A Quantum Source of
  Entropy for Black Holes},'' {\em Phys.Rev.}, vol.~D34, pp.~373--383, 1986.

\bibitem{Jacobson:1995ab}
T.~Jacobson, ``{Thermodynamics of space-time: The Einstein equation of
  state},'' {\em Phys.Rev.Lett.}, vol.~75, pp.~1260--1263, 1995.

\bibitem{Bianchi:2012br}
E.~Bianchi, ``{Horizon entanglement entropy and universality of the graviton
  coupling},'' 2012.

\bibitem{gravastar}
P.~O. Mazur and E.~Mottola, ``{Gravitational condensate stars: An alternative
  to black holes},'' 2001.

\bibitem{zurek}
W.~H. Zurek and D.~N. Page, ``Black-hole thermodynamics and singular solutions
  of the tolman-oppenheimer-volkoff equation,'' {\em Phys. Rev. D}, vol.~29,
  pp.~628--631, Feb 1984.

\bibitem{t'Hooft}
G.~'t~Hooft, ``{The Selfscreening Hawking atmosphere: A New approach to quantum
  black hole microstates},'' {\em Nucl.Phys.Proc.Suppl.}, vol.~68,
  pp.~174--184, 1998.

\bibitem{Afshordi:2006ad}
N.~Afshordi, D.~J. Chung, and G.~Geshnizjani, ``{Cuscuton: A Causal Field
  Theory with an Infinite Speed of Sound},'' {\em Phys.Rev.}, vol.~D75,
  p.~083513, 2007.

\bibitem{Afshordi:2009tt}
N.~Afshordi, ``{Cuscuton and low energy limit of Horava-Lifshitz gravity},''
  {\em Phys.Rev.}, vol.~D80, p.~081502, 2009.

\bibitem{Bredberg:2010ky}
I.~Bredberg, C.~Keeler, V.~Lysov, and A.~Strominger, ``{Wilsonian Approach to
  Fluid/Gravity Duality},'' {\em JHEP}, vol.~1103, p.~141, 2011.

\bibitem{Broderick:2007ek}
A.~E. Broderick and R.~Narayan, ``{Where are all the gravastars? Limits upon
  the gravastar model from accreting black holes},'' {\em Class.Quant.Grav.},
  vol.~24, pp.~659--666, 2007.

\bibitem{Israel}
W.~Israel, ``Singular hypersurfaces and thin shells in general relativity,''
  {\em Il Nuovo Cimento B 19651970}, vol.~571, no.~1964, pp.~1--14, 1966.

\end{thebibliography}

\newpage

\appendix
\section{Israel Junction Condition at $Z_2$ Symmetric Boundary}\label{app1}

In order to get junction conditions at a boundary with radial $Z_2$ symmetry, we will use the same technique as \cite{Israel}.
The metric \eqref{f2} can be written in terms of proper radial distance as
\beq\label{a1}
ds^2=-N^2(\lambda)dt^2+d\lambda^2 +r^2(\lambda)d\Omega^2,
\eeq
where 
\beq
\lambda(r)=\int_{r^*}^r \frac{\mathrm{d}r'}{\sqrt{f(r')}}.\notag
\eeq
Radial $Z_2$ symmetry thus implies that for points at $\lambda=0$, the metric can be locally extended to negative values of $\lambda$, such that 
\bea
N(\lambda)&=&N(-\lambda)\label{a2}\\
r(\lambda)&=&r(-\lambda).\label{a3}
\eea
Also, if there is a thin layer of fluid between $\lambda=0$ and $\lambda=\epsilon$, radial $Z_2$ symmetry will require the existence of a fluid with the same energy momentum tensor between $\lambda=0$ and $\lambda=-\epsilon$.

If $K_{ab}$ and $^3\mathcal{R}_{ab}$ represent the extrinsic curvature and intrinsic curvature of a surface with constant $\lambda$ (which means surface with constant $r$), respectively, then in \eqref{a1} coordinates  \cite{Israel}
\bea
g_{ab}&=&h_{ab}=\text{diag}(-N^2,r^2,r^2\sin \theta)\notag\\
K_{ab}&=&\frac{1}{2}\frac{\partial g_{ab}}{\partial \lambda},\label{app12}\\
\mathcal{R}_{ab}&=-&\frac{\partial K_{ab}}{\partial \lambda}+ Z_{ab}\notag,
\eea
where 
\beq
Z_{ab}=\phantom{}^3\mathcal{R}_{ab}-K K_{ab}+2K^{c}_{a}K_{cb}.\notag
\eeq
Using the Einstein field equation
\beq
R_{\mu \nu}=8 \pi(T_{\mu \nu}-\frac{1}{2}T g_{\mu \nu}),\notag
\eeq
and integrating through the layer (from $\lambda=-\epsilon$ to $\lambda=\epsilon$), we obtain
\beq\label{a4}
K_{ab}(\epsilon)-K_{ab}(-\epsilon)-\int_{-\epsilon}^{\epsilon} \mathrm{d}\lambda Z_{ab}=-8 \pi \int_{-\epsilon}^{\epsilon}\mathrm{d}\lambda (T_{ab}-\frac{1}{2}g_{ab}T).
\eeq
Imposing $Z_2$ symmetry and taking the limit $\epsilon \to 0$ gives
\beq
\mathcal{T}_{ab}\equiv \int_{0}^{\epsilon}\mathrm{d}\lambda T_{ab}=\frac{1}{2}\int_{-\epsilon}^{\epsilon}\mathrm{d}\lambda T_{ab}.\notag
\eeq
Also, \eqref{a2},\eqref{a3} and \eqref{app12} give
\beq
\lim_{\epsilon \to 0}\left[K_{ab}(\epsilon)-K_{ab}(-\epsilon) \right]=2K_{ab} \text{        at $r=r^*$}.\notag
\eeq
In addition, if $K_{ab}$ remains bounded, then
\beq
\lim_{\epsilon \to 0}\int_{-\epsilon}^{\epsilon} \mathrm{d} \lambda \, Z_{ab}=0.\notag
\eeq
In \eqref{a1} coordinates $g_{ab}=h_{ab}$. Consequently, \eqref{a4} results in 
\beq\label{app13}
-8 \pi (\mathcal{T}_{ab}-\frac{1}{2}h_{ab} \mathcal{T})= K_{ab}.
\eeq
The above equation has been derived in a particular coordinate for boundary. However, it is a tensorial equation and accordingly, it is valid in any coordinate for boundary. Taking trace of \eqref{app13} and expressing $T$ in terms of $K\equiv K_{ab}h^{ab}$, we get
\beq
K_{ab}-Kh_{ab}=-8 \pi \mathcal{T}_{ab}\notag
\eeq

We argued that for the special case of spherical symmetry, \eqref{a2} and \eqref{a3} are the conditions required to have $Z_2$ symmetric boundary. This definition can be extended to more general spacetimes.

Let's start with an intuitive definition for $Z_2$ symmetry. A spacetime $(\mathcal{M},g)$ has (local) $Z_2$ symmetry with respect to a hypersurface $S$, which divides spacetime into two parts $(\mathcal{M^+},g^+)$ and $(\mathcal{M^-},g^-)$, if local observers on $S$ cannot distinguish between $\mathcal{M^+}$ and $\mathcal{M^-}$. This means if they move perpendicular to $S$ (along normal vector $n$ to $S$) into $\mathcal{M^+}$ or $\mathcal{M^-}$, they will see the same geometry. In mathematical language, it means
\beq\label{a5}
\mathcal{L}_n\, g^+_{\mu \nu}=\mathcal{L}_{-n} \, g^-_{\mu \nu},
\eeq
where $\mathcal{L}_n$ is Lie derivative with respect to $n$. In particular, it results
\beq
K^+_{ab}=-K^-_{ab},\notag
\eeq
and Israel junction condition \eqref{e3} for $Z_2$ symmetric hypersurface (with space-like normal vector) gives
\beq\label{a6}
2(K_{ab}-K h_{ab})=-8 \pi \mathcal{T}_{ab}.
\eeq

However, if  $S$ is the boundary of spacetime, the situation is a bit different, since $S$ does not divide spacetime into two parts. In this case, $Z_2$ symmetric {\it boundary} means that we glue a copy of spacetime $\mathcal{M}$ to itself through $S$ (it means that $S$ acts like a mirror.) Now, $S$ has divided the whole spacetime ($\mathcal{M}+\mathcal{M}$) into two parts and condition \eqref{a5} has been satisfied. However, we must multiply the right hand side of \eqref{a6} by a factor of two, because there is also a copy of the surface fluid on the other side (as we showed concretely for spherically symmetric spacetimes.)

As a result, we obtain
\beq\label{a7}
K_{ab}-K h_{ab}=-8 \pi \mathcal{T}_{ab},
\eeq
for $Z_2$ symmetric boundaries. Indeed, equation \eqref{a7} can be used as a definition for $Z_2$ symmetric boundary.
\section{Near Horizon Geometry of Kerr-Newman Black Hole}\label{app2}

The Kerr-Newman metric describes the geometry of a black hole with angular momentum $J$ and charge $Q$. This metric can be written as
\bea
ds^2&=&g_{tt}dt^2+g_{rr}dr^2+g_{\theta \theta}d\theta^2+g_{\phi \phi}d\phi^2+2 g_{t \phi}dt d\phi\notag\\
&=& (g_{tt}-\frac{g_{t \phi}^2}{g_{\phi \phi}})dt^2+g_{rr}dr^2+g_{\theta \theta}d\theta^2+g_{\phi \phi}(d\phi+\frac{g_{t \phi}}{g_{\phi \phi}}dt)^2,\notag
\eea
where
\bea
g_{tt}&=&-(1-\frac{2m r-r_Q^2}{\Gamma}),\notag\\
g_{rr}&=& \frac{\Gamma}{\Delta},\notag\\
g_{\theta \theta}&=&\Gamma,\notag\\
g_{\phi \phi}&=&\sin^2 \theta\left(r^2+a^2+\frac{(2mr-r_Q^2)a^2\sin^2\theta}{\Gamma}\right),\notag\\
g_{t \phi}&=&-\frac{(2mr -r_Q^2)a\sin^2\theta}{\Gamma},\notag
\eea
with
\bea
\Delta&=&r^2-2mr+a^2+r_Q^2,\notag\\
\Gamma&=&r^2+a^2\cos^2\theta,\notag
\eea
and $a=\frac{J}{m}$ and $r_Q=\frac{Q^2}{4 \pi \epsilon_0}$.
The horizon of Kerr-Newman metric is at $\Delta=0$
\beq
r^2-2mr+a^2+r_Q^2=0 \rightarrow r_{\pm}=m\pm \sqrt{m^2-a^2-r_Q^2}.\notag
\eeq

In order to derive the near horizon geometry of the Kerr-Newman metric we   define a new variable
\beq
\lambda=\int_{r_{+}}^r \frac{dr'}{\Delta(r')}\approx 2\sqrt{\frac{r-r_+}{r_+-r_-}}+O((r-r_+)^{3/2}).\notag
\eeq
Small values of $\lambda$ correspond to radii close to the horizon $r_+$. Replacing $r$ in the metric with the new coordinate $\lambda$ and keeping the leading terms for $\lambda \ll 1$, we obtain
\bea
\Delta&=& (r-r_+)(r-r_-)\approx \frac{(r_+-r_-)^2}{4}\lambda^2,\notag\\
\Gamma&=&r^2+a^2\cos^2 \theta \approx r_+^2+a^2 \cos^2 \theta \equiv \Gamma_+,\notag\\
g_{tt}-\frac{g_{t \phi}^2}{g_{\phi \phi}}&\approx& -\frac{(r_+-r_-)^2}{4}\frac{\Gamma_+\, \lambda^2}{(r_++a^2)^2},\notag\\
g_{rr}&\approx& \frac{4 \Gamma_+}{(r_+-r_-)^2\lambda^2},\notag\\
g_{\theta \theta}&\approx &\Gamma_+,\notag\\
\frac{g_{t \phi}}{g_{\phi \phi}}&\approx& -\frac{a}{r_+^2+a^2}\equiv -\Omega,\notag\\
g_{\phi \phi}&\approx& \frac{(r_+^2+a^2)^2\sin^2 \theta}{\Gamma_+}.\notag
\eea
Defining new variables $\psi\equiv (\phi-\Omega t)(r_+^2+a^2)$ and $\tau \equiv \frac{r_+-r_-}{2(r_+^2+a^2)}t$, we get
\beq\label{ap2}
ds^2=-\Gamma_+ \lambda^2 d\tau^2+\Gamma_+d\lambda^2+\Gamma_+d\theta^2+\frac{\sin^2\theta}{\Gamma_+}d\psi^2.
\eeq





\end{document}